\documentclass[aps,pre,superscriptaddress,twocolumn]{revtex4}

\usepackage{graphicx}
\usepackage{amsfonts}
\usepackage{amssymb}
\usepackage{amsmath}
\usepackage{color,verbatim} 
\usepackage[pdftex]{hyperref}
\hypersetup{colorlinks=true,linkcolor=blue,citecolor=blue,urlcolor=blue}

\begin{document}

\title{Stability and dynamics of massive vortices in two-component 
Bose-Einstein condensates}

\author{J. D'Ambroise}
\affiliation{Department of Mathematics, 
Computer $\&$ Information Science, 
State University of New York (SUNY) College at Old Westbury, 
Westbury, NY, 11568, USA}

\author{W. Wang}
\affiliation{College of Physics, Sichuan University, 
Chengdu 610065, China}

\author{C. Ticknor}
\affiliation{Theoretical Division, Los Alamos National Laboratory, 
Los Alamos, New Mexico 87545, USA}

\author{R. Carretero-Gonz{\'a}lez}
\affiliation{Nonlinear Dynamical Systems Group,
Computational Sciences Research Center, and
Department of Mathematics and Statistics,
San Diego State University, San Diego, California 92182-7720, USA}

\author{P.G. Kevrekidis}
\affiliation{Department of Mathematics and Statistics, 
University of Massachusetts, Amherst, MA 01003-4515, USA}

\begin{abstract}
The study of structures involving vortices in one component
and bright solitary waves in another has a time-honored history
in two-component atomic Bose-Einstein condensates.
In the present work, we revisit this topic extending considerations
well-past the near-integrable regime of nearly equal scattering lengths.
Instead, we focus on stationary states and spectral stability of
such structures for large values of the inter-component interaction
coefficient. We find that the state can manifest 
dynamical instabilities for suitable parameter values.
We also explore a phenomenological, yet quantitatively accurate
upon suitable tuning, particle model which, in line also with 
earlier works, offers the potential of accurately following the
associated stability and dynamical features. 
Finally, we probe the dynamics of the unstable vortex-bright 
structure, observing an unprecedented, to our knowledge, 
instability scenario in which the oscillatory instability
leads to a patch of vorticity that harbors and eventually ejects
multiple vortex-bright structures.
\end{abstract}

\maketitle

\section{Introduction}
\label{sec:intro}

The platform of atomic Bose-Einstein condensates (BECs) has been 
particularly fruitful towards the exploration of nonlinear
wave patterns, both solitonic ones, as well as vortical ones involving
topological charge~\cite{pethick,stringari}. The ability to
manipulate external potentials in both space and time~\cite{boshier},
as well as that of controlling the size and the sign of the
atomic scattering length (and hence the coefficient of nonlinear
interaction) in time~\cite{Donley_2001,hulet1} or
even in space~\cite{borisrmp,yiota} have paved new directions
of unprecedented perturbations and dynamics to the solitary
wave patterns that such systems can support.

In addition to the spatio-temporal control and the wealth of
phenomenology of one-component atomic systems, 
two-component ones~\cite{siambook,DJF2C} are, arguably, 
even richer in their potential phenomenology. On the other hand,
the exploration of spinor systems~\cite{kawaguchi2012spinor,stamper2013spinor} has enabled
the consideration of magnetic, as well as non-magnetic 
structures~\cite{li2005exact,zhang2007solitons,nistazakis2008bright,szankowski2011surprising,romero2019controlled,chai2020magnetic,chai2021magnetic,bersano2018three,fujimoto2019flemish}.
These also include spin domains~\cite{miesner1999observation,swislocki2012controlled}
and spin textures~\cite{ohmi1998bose,song2013ground}, taking advantage
of the spin degree of freedom.
In this setting, the ground state has been recognized to have the
potential
for antiferromagnetism in $^{23}$Na or ferromagnetism in $^{87}$Rb.
However, perhaps even more interestingly, the wealth of such systems
lies in the excited and topological structures that they can support
in  the
three-component $F\!=\!1$ and the five-component $F\!=\!2$ settings~\cite{kawaguchi2012spinor,stamper2013spinor}.

While these higher-component settings offer considerable
additional wealth in terms of the possible states, 
our aim herein is to revisit two-component settings. 
There, many solitary waveforms have been considered, 
ranging from dark-bright to dark-dark  and
dark-antidark, as well as magnetic solitons~\cite{siambook,DJF2C,PhysRevLett.116.160402}. 
Not only individual structures of this kind have been experimentally 
explored, but also small clusters thereof (see, e.g.,
Ref.~\cite{Katsimiga2020}), as well as their collisions~\cite{becker2008}.
Such two-component settings have recently been leveraged
also towards engineering effective attractive interactions
in both 1D and 2D, offering the possibility to generate
structures such as the Townes soliton~\cite{dalibard},
as well as the Peregrine soliton~\cite{peregrine}.

Here, we focus on the 2D
variants of the above solitary waves in the form of vortex-bright (VB)
solitons~\cite{skryabin,tuckerman}. These are sometimes referred to by other
names such as baby-skyrmions, filled-core vortices, or massive vortices 
and have not only been theoretically explored but also experimentally 
observed~\cite{bpa1,wolf,yongil}. Note that some of these experimental
realizations such as the one reported in Ref.~\cite{yongil} 
(and corresponding theoretical explorations such as in Ref.~\cite{liavb}) 
have probed these states in spinor settings.
In recent years there has been a flurry of activity in such systems,
seeking to provide a theoretical understanding of the potential
dynamics of these structures. Indeed, they have been shown 
to be able to perform flower-like motions~\cite{RichaudFetter:21,richaudannulus}; 
they have been also explored in generalized radial potentials~\cite{Wang:VB,richaudgeneral},
as well as in the context of rotating traps~\cite{richaudmuntsa}
and more recently the case of multiple such states and how
they interact has been revisited~\cite{richaudmultiple}
(see also earlier studies in Refs.~\cite{PSHENICHNYUK2018523,nitta}).
It is relevant to point out here that in the context
of miscible BECs the interaction of these VB solitons was 
considered earlier in Ref.~\cite{etonitta}. An additional,
intriguing in its own right, setting where a superfluid bearing
a vortex is coupled to an impurity, i.e., a second component
without self-interactions but solely feeling the vortex as
an ``external potential'' was recently proposed in Ref.~\cite{prr}.
Both the VB-like state but also higher excited states of the
second component impurity were considered therein. Finally, 
yet another direction touched upon recently in the VB
setting is that of higher topological charges, where it was
found that the bright component could play a significant
role in their dynamical stabilization~\cite{barenghi}.

Here, our aim is to complement some of these developments, which
have chiefly been based on Lagrangian (variational) analytical
formulations and direct numerical simulations, by means of steady 
state computations, Bogoliubov-de Gennes (BdG) stability
analysis, dynamics, and a modified ODE analytical framework. In Sec.~\ref{sec:thy}, we provide a palette of relevant tools
upon discussing the model. Notice that here we are primarily 
focusing on scenarios that are deep within the immiscible regime,
in line with some of the above works, e.g., by Richaud and
collaborators~\cite{RichaudFetter:21,richaudannulus,richaudgeneral,richaudmultiple}. We also discuss the details of an analytical approach that
has some modified features in comparison to the above works. 
We analyze the types of motions that are permissible in this
reduced system and provide their quantitative characteristics
towards a comparison with the full numerical PDE results.
In Sec.~\ref{sec:num}, we provide the numerical continuation of our VB
states as a function of the system's chemical potentials for different values of the inter-species coupling. Here, we also offer a systematic procedure for matching the ODE and PDE dynamics,
and showcase that this matching is also reflected in the BdG spectrum
of the VB matching the corresponding particle
model eigenfrequencies. Moreover, we trace an (oscillatory) 
instability of the
PDE VB state when two pertinent eigenfrequencies collide and become complex.
These findings, in turn, prompt us to perform select dynamics,
reflecting some of the more elaborate VB features, such as the
flower-like trajectories thereof, but also the instabilities of such
structures. The latter turn out to be especially intriguing as the
oscillatory instability results in the formation of a patch of vorticity
harboring multiple VB structures that eventually lead to a breakup
of the relevant pattern in different segments. Finally, in Sec.~\ref{summary},
we summarize our findings and present our conclusions.

\section{Model and Theoretical Setup}
\label{sec:thy}
\subsection{Full PDE model}
\label{sec:PDE}

To study the existence, stability, and dynamics of VB complexes,
let us consider the 2D Gross-Pitaevskii (GP) equation
describing the interaction between two different atomic species in a BEC.
The corresponding GP equation can be cast in non-dimensional form as:
\begin{subequations}
\label{eq:GPE}
\begin{eqnarray}
i \dot\Psi_1&=&\displaystyle-\frac{1}{2} \Delta \Psi_1+\left[ V(r)+ |\Psi_1 |^2 + g_{12}|\Psi_2|^2\right] \Psi_1,
~~~~
\label{eq:GPEa}
\\[1.0ex]
\label{eq:GPEb}
i \dot\Psi_2&=&\displaystyle-\frac{1}{2} \Delta \Psi_2+\left[ V(r) + |\Psi_2 |^2 + g_{12}|\Psi_1|^2\right] \Psi_2,
~~~~
\end{eqnarray}
\end{subequations}
where both BEC wavefunction species $\Psi_1$ and $\Psi_2$ are confined
by the common external parabolic potential $V(r) = \frac{1}{2}\Omega^2r^2$ with strength
$\Omega$ which is usually deployed in the experiments using a magnetic field~\cite{stringari,pethick,siambook}. 
Here, $\Delta$ denotes the 2D Laplacian, the 
overdot is used for differentiation with 
respect to time, and $g_{12}>0$ is the adimensional
inter-species coupling.
It is relevant to point out here that we have
in mind a setup such as $^{87}$Rb (or similar)
in which the intra-species scattering lengths
are practically equal, while we think of the
inter-species one as being tunable via external
fields~\cite{stringari,siambook}.
To construct VB complexes we consider the first component
to be populated with a relatively large number of atoms so that it is 
close to its Thomas-Fermi (TF) limit. We will dub this component the 
``dark'' component as it will be primarily governed by the self-repulsive
interaction present in the first component [cf.~positive sign in front
of $|\Psi_1|^2$ in Eq.~(\ref{eq:GPEa})] and it
will be featuring a vortex structure.
In turn, the vortex embedded in the dark component will induce an
effective potential on the second component through the term
$g_{12}|\Psi_1|^2 \Psi_2$ in Eq.~(\ref{eq:GPEb}) that will trap 
a relatively small number of atoms and create a localized hump 
that we will dub the ``bright'' component. 
This is the origin of the terminology associated
with the vortex-bright 
(or dark-bright in 1D~\cite{DJF2C}) complex.
Solutions to Eq.~(\ref{eq:GPE}) conserve the atom number (power) 
in each of the two components:
\begin{equation}
P_j = \iint_{-\infty}^{+\infty} |\psi_j|^2 \,dx\, dy
\quad{\rm for}\quad j=1,2.  
\label{eq:power}
\end{equation}
An additional important parameter for the VB complex is the ---also conserved--- mass ratio:
\begin{equation}
M = \frac{P_2}{P_1}. 
\label{eq:M}
\end{equation}

Let us now seek stationary state solutions of the form 
$\Psi_1(x,y,t) = \psi_1(x,y)\,e^{-i\mu_d t}$ and  
$\Psi_2(x,y,t) = \psi_2(x,y)\,e^{-i\mu_b t}$  
where $\mu_d$ and $\mu_b$ denote the chemical potentials of the dark 
and bright components, respectively. 
Then, the spatial part of the stationary solutions satisfies
\begin{subequations}
\label{eq:stat}
\begin{eqnarray}
-\frac{1}{2} \Delta \psi_1+\left[ V(r)+  |\psi_1 |^2 + g_{12}|\psi_2|^2
-\mu_d\right] \psi_1&=&0,
~~~~
\label{eq:stata}
\\[1.0ex]
\label{eq:statb}
-\frac{1}{2} \Delta \psi_2+\left[ V(r) +  |\psi_2 |^2 + g_{12}|\psi_1|^2
-\mu_b\right] \psi_2&=&0.
~~~~
\end{eqnarray}
\end{subequations}
The VB states are studied in the relevant parameter space. Here, the key parameters determining the properties of the VB complex are $\Omega$, $g_{12}$, $\mu_d$, and $\mu_b$. For example, the powers and the mass ratio depend directly, in a non-trivial manner, on the chemical potentials for a given set of $\Omega$ and $g_{12}$ values.

We find numerically exact stationary states using finite 
differencing for the spatial discretization together with 
a Newton-based fixed-point iteration method. 
First, we find a relevant solution, and then numerical continuation can be deployed along any of the above parameters. The first solution is obtained by embedding a vortex centered at the origin in the first component, we shall discuss the details in Sec.~\ref{sec:num}. We also study the (spectral) stability of these centered VB complexes by the BdG analysis. Then, taking parameter values where the VB complex is stable, we proceed to appropriately seed it away from the trap center and monitor, according to 
Eq.~(\ref{eq:GPE}), its dynamical evolution. Our dynamics is carried out using the regular fourth-order Runge-Kutta method. 
The location of the VB complex is monitored
by first computing the respective centers of mass of each component
along the $x$ and $y$ directions:
\begin{equation}
{\rm CoM}^\sigma_j =\frac{1}{P_j} \iint_{-\infty}^{+\infty} \sigma|\Psi_j|^2 \,dx\,dy,
\label{eq:CoM} 
\end{equation}
where $\sigma=x,y$. Then, the relative location of the VB with 
respect to the background cloud of the dark component is recorded 
as ${\rm CoM}^\sigma_2-{\rm CoM}^\sigma_1$. 

The BdG analysis for the two-component stationary solutions works 
as follows. Given the steady state solutions $(\psi_1, \psi_2)$ 
as per Eq.~(\ref{eq:stat}), we perturb them according to
\begin{eqnarray}
  \Psi_1 &=& \left\{\psi_1 + \varepsilon\left[a_1(x,y)e^{\nu t} + b_1^*(x,y) e^{\nu^* t}\right]\right\}e^{-i\mu_d t}, 
 ~~~
\notag\\
&&  \label{eq:perturb}\\
  \Psi_2 &=& \left\{\psi_2 + \varepsilon\left[a_2(x,y)e^{\nu t} + b_2^*(x,y) e^{\nu^* t}\right]\right\}e^{-i\mu_b t}.  
 ~~~
\notag
\end{eqnarray}
The spectra are then computed by plugging Eq.~(\ref{eq:perturb}) 
into Eq.~(\ref{eq:GPE}) and solving the resulting linearized 
eigenvalue problem at order $\varepsilon$ as:
\begin{equation}
\notag
-i\left[ \begin{array}{cccc} 
\phantom{-}M_{11} & \phantom{-}M_{12} & \phantom{-}M_{13} & \phantom{-}M_{14} \\ 
-M_{12}^* & -M_{11}^* & -M_{14}^* & -M_{13}^* \\ 
\phantom{-}M_{13}^* & \phantom{-}M_{14} & \phantom{-}M_{22} & \phantom{-}M_{34} \\ 
-M_{14}^* & -M_{13} & -M_{34}^* & -M_{22}^* \\ 
\end{array} \right] 
\left[ \begin{array}{c} a_1\\ b_1\\ a_2 \\ b_2 \end{array} \right] 
=  \nu 
\left[ \begin{array}{c} a_1\\ b_1\\ a_2 \\ b_2 \end{array} \right] ,
\end{equation}
where
\begin{eqnarray}
M_{11} &\equiv& -\frac{1}{2} \Delta + V- \mu_d + 2|\psi_1|^2 + g_{12}|\psi_2|^2, \notag\\[1.0ex]
M_{22} &\equiv& -\frac{1}{2} \Delta + V- \mu_b + 2|\psi_2|^2 + 
g_{12}|\psi_1|^2, \notag
\end{eqnarray}
and 
$M_{12} \equiv \psi_1^2$, 
$M_{13} \equiv g_{12}\psi_1\psi_2^*$, 
$M_{14} \equiv g_{12}\psi_1\psi_2$, and 
$M_{34} \equiv \psi_2^2$.   
The eigenvector for the corresponding eigenvalue $\nu$ is $[\Psi_1,\Psi_2]^T=[(a_1,b_1),(a_2,b_2)]^T$, and the mode is unstable if Re$(\nu)>0$. 
As we will see in Sec.~\ref{sec:num}, after stationary solutions are 
displaced away from the origin and for appropriate parameter regimes, the VB states can exhibit 
flowering trajectories. 
%
To follow potential instabilities emerging from the
collision of eigenvalues, for each eigenvector we monitor 
the sign of the quantity (so-called Krein signature)~\cite{kks}:
\begin{equation}
K = \iint_{-\infty}^{+\infty} (|a_1|^2-|b_1|^2+|a_2|^2-|b_2|^2)\,dx\,dy.
\notag
\end{equation}
Collision of two dynamically stable [Re$(\nu)=0$] eigenvalues of opposite 
Krein sign generically leads to an oscillatory instability~\cite{skryabin}.
In Sec.~\ref{sec:num} we detail  the presence of
a supercritical Hamiltonian Hopf bifurcation when continuing
along the $\mu_b$ parameter. This is induced precisely by 
the collision of two opposite Krein sign
eigenvalues that correspond to the
observed flowering frequencies of the VB evolution;
see more details on that below.

\subsection{Reduced ODE model for the VB complex}
\label{sec:ODE}

We now turn to a theoretical description of the particle motion
of the VB structure inside the parabolic trap. This will provide
a yardstick of comparison for the parametric dependence of
the PDE results that will be obtained below based on the methods
presented in Sec.~\ref{sec:PDE}. 
Our model somewhat differs from the one used, e.g., 
in Ref.~\cite{RichaudFetter:21} in the following two ways.
The latter model contains a singular logarithmic
term in its potential energy, while the model used herein
contains an effective potential energy essentially proportional
to the system density at the location of interest (see details
below). Furthermore, the term giving rise to the
Magnus force and being effectively associated
with the angular momentum also emerges linearly 
in its dependence on the density (leading to the right equation
in the single-component vortex limit).  In that vein,
based on the above motivations,
we follow the approach of Ref.~\cite{ruban22}, using effectively
as the dynamical model of relevance the one of Eq.~(29) therein.

The ODE model can be reduced to the following second order system
on the VB position $(x(t),y(t))$:
\begin{eqnarray}
M_0\,\ddot{x} &=& \frac{\Lambda}{2}x - \rho(r)\dot{y}\notag\\
\ &&  \label{eq:ode}\\
M_0\,\ddot{y} &=& \frac{\Lambda}{2}y + \rho(r)\dot{x}\notag
\end{eqnarray}
where $M_0$ and $\Lambda$ depend, in a non-trivial manner, on the 
relative mass $M$ and the inter-species coupling $g_{12}$~\cite{ruban22},  
and $\rho(r) \equiv \mu_d - V(r)$ 
is the TF background density of the dark component.   
The equations of motion can be recovered through the Euler-Lagrange 
equations  $\frac{d}{dt}\left( \frac{\partial L}{\partial \dot{\sigma}}\right) 
= \frac{\partial L}{\partial \sigma}$, for $\sigma=x,y$,
from the corresponding Lagrangian:
\begin{equation}
L = \frac{M_0}{2}(\dot{x}^2+\dot{y}^2) - \frac{1}{2}(x\dot{y} - y\dot{x})\left( \mu_d-\frac{1}{2}V(r)\right)  + \frac{\Lambda}{4}r^2.
\label{eq:L}
\end{equation}
Note that we may also write $L$ in polar coordinates using standard 
conversion formulas $x^2+y^2=r^2$, $\dot{x}^2+\dot{y}^2 
= \dot{r}^2+r^2\dot\theta^2$, and $x\dot{y} - y\dot{x} = r^2\dot\theta$ 
for $r=r(t)$ and $\theta=\theta(t)$.   
Comparing Eq.~(\ref{eq:L}) above with the Lagrangian~(23) of 
Ref.~\cite{RichaudFetter:21}, we find similar structural features 
with the expression used herein bearing an additional contribution  $\left[ \mu_d-\frac{1}{2}V(r)\right]$ 
in the second kinetic energy term, while instead of the logarithmic term in Ref.~\cite{RichaudFetter:21}, 
we include herein only the quadratic term in its Taylor series.  
Additionally, their $r$ is normalized as $r/R_{\rm TF}$ for $R_{\rm TF}$ 
the TF radius of the dark component.

The conservation of angular momentum in polar coordinates 
can be inferred from 
the angular Euler-Lagrange equation $\frac{d}{dt}\left( \frac{\partial L}{\partial 
\dot{\theta}}\right) = \frac{\partial L}{\partial \theta}$ 
which yields:
\begin{equation}
\ell = \Omega^2M_0 r^2\dot{\theta} + \frac{1}{2}\rho^2(r).
\label{eq:angmom}
\end{equation}
On the other hand, to derive the Hamiltonian formulation of the 
ODE system~(\ref{eq:ode}) we can utilize
the Legendre transform $H = \dot{x}p_x  
+ \dot{y} p_y - L$ with $p_\sigma=\frac{\partial L}{\partial \dot{\sigma}}$ for $\sigma=x,y$, to obtain 
\begin{eqnarray}
H &=& \frac{1}{2M_0}\left(p_x^2+p_y^2 - (yp_x-xp_y)\left(\mu_d-\frac{1}{2}V(r)\right) \right) \notag\\
&&+ \frac{r^2}{8M_0}\left(\mu_d-\frac{1}{2}V(r)\right)^2-\frac{\Lambda}{4}r^2.
\notag
\end{eqnarray}
The effective total (and conserved) energy can then be written
as follows:
\begin{equation}
E = \frac{1}{2}M_0\dot{r}^2 + V_{\rm eff}(r),
\label{eq:E}
\end{equation}
with the effective potential 
\begin{equation}
V_{\rm eff}(r) = \frac{1}{2}M_0r^2\dot{\theta}^2 - \frac{\Lambda}{4}r^2.
\notag
\end{equation}
Using Eq.~(\ref{eq:angmom}) to replace $\dot{\theta}^2$ with the 
corresponding expression in terms of $\ell$ we can also write the 
effective potential as:
\begin{equation}
V_{\rm eff}(r) = \frac{1}{2\Omega^4 M_0r^2}\left(\ell -  \frac{1}{2} \rho(r)^2 \right)^2 - \frac{\Lambda}{4}r^2. 
\label{eq:Veffell}
\end{equation}

The Euler-Lagrange equation in the radial variable 
$\frac{d}{dt}\left( \frac{\partial L}{\partial \dot{r}}\right) = \frac{\partial L}{\partial r}$ yields:
\begin{equation}
M_0(\ddot{r} - r\dot\theta^2) = -\rho(r)r\dot\theta + \frac{\Lambda}{2}r,
\notag
\end{equation}
which can be used to derive the expected precessional frequency of 
a circular trajectory by setting $r=r_0$ and $\ddot{r}=0$ to obtain 
the quadratic-in-$\dot\theta$ equation
\begin{equation}
0 = M_0\dot\theta^2 -  \rho(r_0)\dot\theta + \frac{\Lambda}{2}.
\label{eq:quadratic}
\end{equation}
Solving for $\dot\theta$ yields frequencies
\begin{equation}
\dot\theta_{\pm} = \frac{1}{2M_0}\left(\rho(r_0)\pm\sqrt{\rho(r_0)^2-2M_0\Lambda}\right).
\label{eq:freqpm}
\end{equation}
The smaller frequency $\omega_{\rm pr}\equiv\dot\theta_{-}$ corresponds to the expected precessional frequency of a circle, 
and reduces to the expected single-component
precession frequency asymptotically in the limit of
$M_0 \rightarrow 0$.
Here, given the presence of the second component, 
we also obtain a larger frequency $\dot\theta_{+}$ that 
also pertains to a precessional motion, but indeed one whose
frequency diverges in the $M_0 \rightarrow 0$ limit.

For simplicity, we assume that a circle begins with $x(0)=x_0$, $y(0)=0$,
and $\dot{x}(0)=0$. This yields, according to Eq.~(\ref{eq:quadratic}) 
converted to Cartesian coordinates, the initial velocity for a circle 
in the ODE model as: 
\begin{equation}
\dot{y}(0) =\frac{1}{2M_0}\left( \rho(r_0)x_0\pm\sqrt{\rho(r_0)^2 x_0^2 - 2M_0\Lambda r_0^2}\right).
\notag
\end{equation}
Finally, we note that the resulting frequencies of flowering 
trajectories can be estimated by converting the conservation 
Eqs.~(\ref{eq:angmom})--(\ref{eq:E}) into a first order system 
and linearizing as follows. 
First off, the potential flowering pertains to small
oscillations in the radial direction
around an equilibrium circular trajectory $r=r_0$.
Differentiating Eq.~(\ref{eq:E}) and solving for $\dot\theta$ in 
Eq.~(\ref{eq:angmom})  we can write the system as $\dot{r}=u$,
$\dot{u}=\ddot{r}=G(r)$, and $\dot\theta=K(r)$ for 
\begin{eqnarray}
G(r) &\equiv& -\frac{1}{M_0}V'_{\rm eff}(r)\notag\\[1.0ex]
K(r) &\equiv& \frac{1}{\Omega^2 M_0 r^2}\left( \ell - \frac{1}{2}\rho(r)^2 \right).\notag
\end{eqnarray}
Near the minimum $r_0$ of $V(r)$ we estimate
\begin{eqnarray}
r(t)&\approx& r_0 + R\cos(\omega_r t)\notag\\
&&\label{eq:estRTheta}\\
\theta(t) &=& \omega_{\rm pr} t + A_0\sin(\omega_r t)\notag
\end{eqnarray}
for radial and precessional frequencies $(\omega_r, \omega_{\rm pr})$
and $R$ and $A_0$ being small perturbations off of the circular motion. 
Plugging Eqs.~(\ref{eq:estRTheta}) into the first order system above gives 
\begin{eqnarray}
\omega_r^2 &=& \frac{V''_{\rm eff}(r_0)}{M_0},\notag\\
&&
\label{eq:omegaA}
\\
A_0 &=& \frac{R K'(r_0)}{\omega_r},\notag
\end{eqnarray}
for $R=r(0)-r_0$.  Note that, according to Eqs.~(\ref{eq:estRTheta}), 
flowering trajectories with looping petals occur when 
${\rm min}\{\dot\theta(t)\}=\omega_{\rm pr}-\omega_r A_0$ is negative 
so that looping appears at the threshold amplitude 
$A_0=\omega_{\rm pr}/\omega_r$ according to the linearized prediction.

In the next section, we will make the correspondence between the PDE 
model and its effective ODE reduction using the following program.  
Since we expect frequencies $\dot\theta$ from the ODE model 
to correspond to purely imaginary frequencies in the PDE model, 
we use the BdG bifurcating eigenvalues $\nu$ 
and the correspondence $\dot\theta = {\rm Im}(\nu)$
in order to estimate the ODE 
parameter value of $\Lambda$ in terms of $M_0$.  
The value of $M_0$ is then determined by minimizing the sum of 
squares of differences between the PDE and ODE trajectories.  
For a given pair of $M_0$ and $\Lambda$ one then finds
initial velocities of the ODE model that best align to a 
given PDE trajectory.  In general, a larger initial velocity gives 
flowering trajectories with more looping on the petals so this can 
be used to guide best choices of these initial velocities in the ODE.  
For the obtained flowering trajectories, we will also show examples 
comparing the ODE trajectories according to Eq.~(\ref{eq:ode}) 
versus the results of the linearized predictions in 
Eqs.~(\ref{eq:estRTheta}) and (\ref{eq:omegaA}).

\section{Results}
\label{sec:num}
\subsection{Steady states and stability for the PDE model}
\label{sec:GPESS}

To obtain the stationary solutions of Eq.~(\ref{eq:stat}), the ground 
state cloud $(\psi_1,\psi_2)=(\psi_{\rm gs},0)$ for the isolated
dark component is found via a standard fixed point Newton's iteration
method with an initial guess corresponding to the TF approximation 
$\psi_1 = \sqrt{\max(\mu_d-V(r),0)}$.   
In turn, to find a VB solution of Eq.~(\ref{eq:stat}) centered
at the origin, we again use  Newton's method, this time with the initial 
guess in the bright component as $\psi_2=\sqrt{\mu_b}\, \rm sech(\mu_b r)$ 
and in the dark component as 
$\psi_1= \psi_{\rm gs}\tanh(\mu_d r)e^{-i\theta(x,y)}$ 
where $\theta$ is the (polar) angle of $(x,y)$ from the origin.  
The resulting unit (negative) charge VB stationary centered 
solutions exist for a variety of combinations of the parameters 
$\Omega$, $g_{12}$, $\mu_b$, and $\mu_d$.  
Once a relevant solution is found, it can be further continued 
along any of these parameters. Similarly, if we take the 
complex conjugate of the initial guess, we can obtain a VB state 
with an opposite charged vortex.

\begin{figure}
\includegraphics[width=1.0\columnwidth]{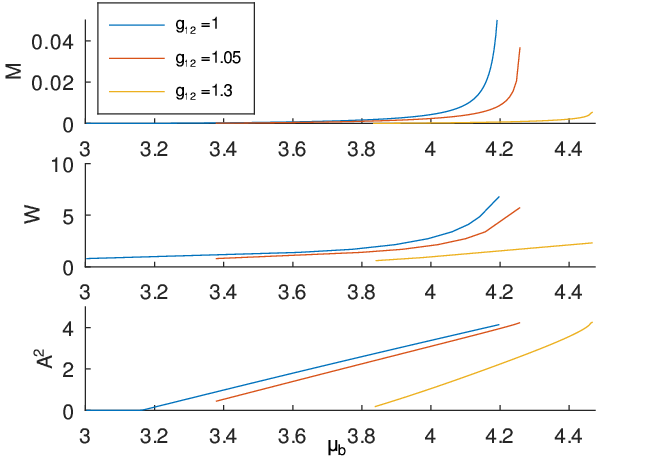}
\caption{
Static properties of the VB steady state centered at
the origin. From top to bottom: mass ratio $M$, full width at half max $W$,
and squared amplitude $A^2$ of the bright component.
These properties are plotted for relatively small values of the inter-species coupling $g_{12}$ for fixed values of
$\mu_d=4.22$ and $\Omega=0.1$.
For these relatively small values of $g_{12}$, all properties
increase with increasing $g_{12}$.
}
\label{fig:statsolpropabc}
\end{figure}

\begin{figure}
\includegraphics[width=1.0\columnwidth]{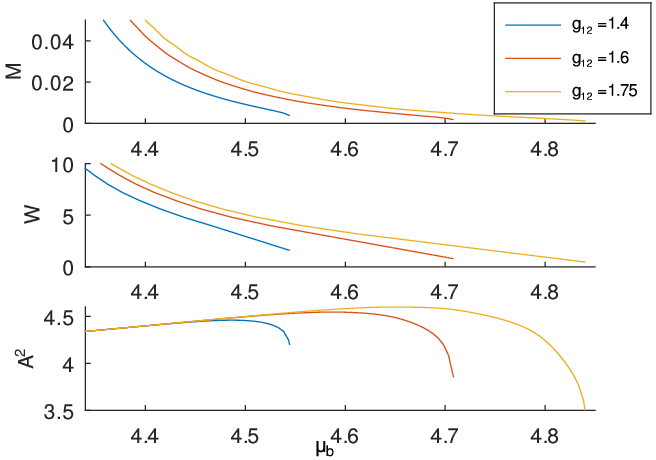}
\caption{
Same as Fig.~\ref{fig:statsolpropabc} but for higher $g_{12}$ values. 
}
\label{fig:statsolpropdef}
\end{figure}

\begin{figure}
\includegraphics[width=0.9\columnwidth]{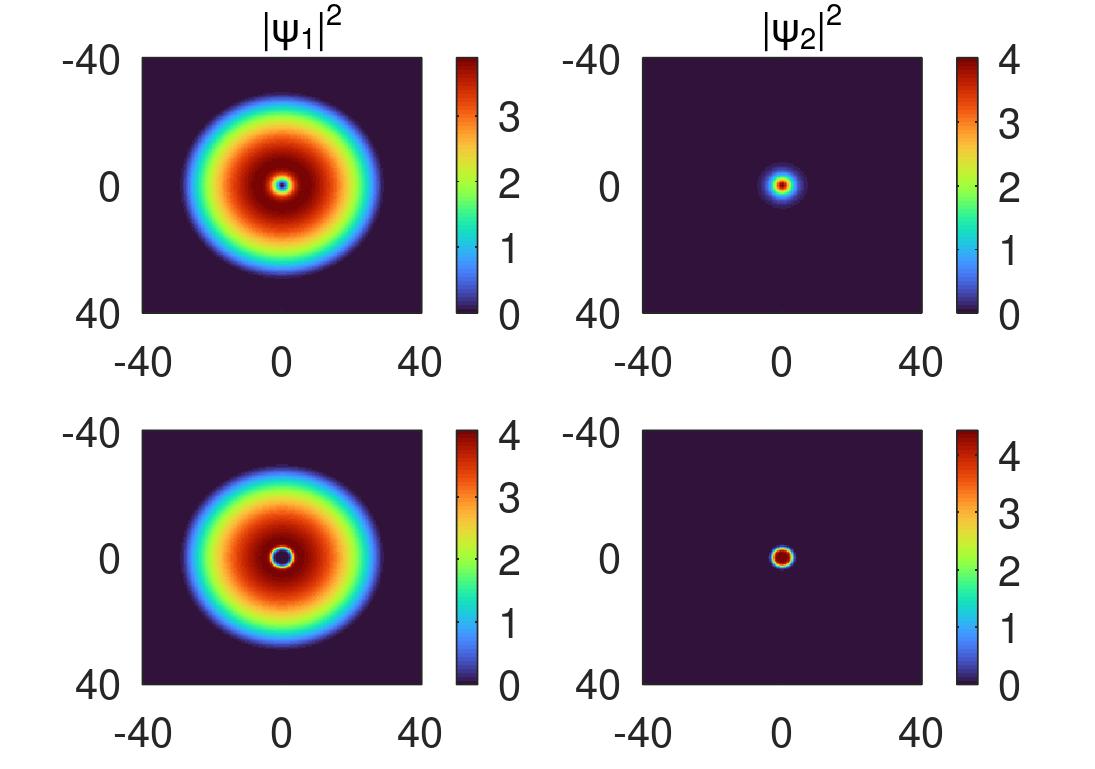}
\caption{
Vortex-bright complexes. Depicted are the densities of two steady state 
solutions with parameter values $\Omega=0.1$ and $\mu_d=4.22$ where 
both solutions have mass ratio $M\approx 0.02$.   
The top row solution corresponds to $(g_{12},\mu_b)=(1,4.16)$ and 
the solution appears to have soft edges with the bright spot having 
a wider footprint.  
By contrast, the bottom row solution corresponds to  
$(g_{12},\mu_b)=(1.4,4.43)$ where the solution shows a sharper boundary 
with the bright spot featuring a slightly narrower footprint.  
}
\label{fig:statsol-imagesc}
\end{figure}

\begin{figure}
\includegraphics[width=1.0\columnwidth]{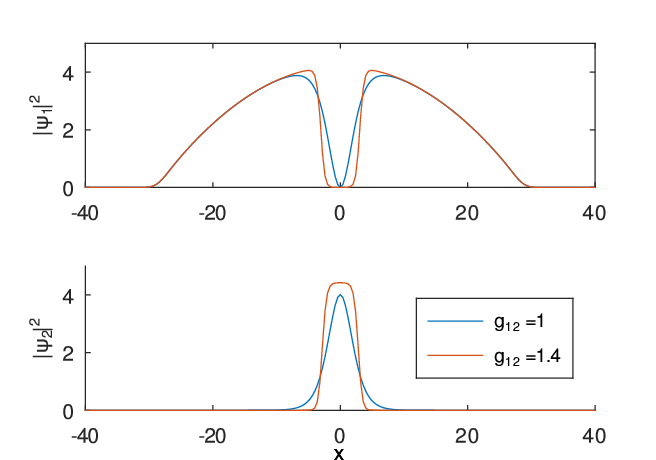}
\caption{
Cross sections of the solutions depicted in Fig.~\ref{fig:statsol-imagesc}.
The top (bottom) panel illustrates the dark (bright) component for the
values of $g_{12}$ indicated.
}
\label{fig:statsol-cut}
\end{figure}

We monitor three properties of the stationary VB states: the full width half max (FWHM) $W$, the squared amplitude $A^2$ of the bright 
component, and the mass ratio $M$ of the two components as defined 
in Eq.~(\ref{eq:M}).
These properties are computed along a continuation in $\mu_b$ for 
fixed values of $g_{12}$, $\mu_d$, and $\Omega$, and they are depicted in
Figs.~\ref{fig:statsolpropabc} and \ref{fig:statsolpropdef}.  
For the relatively small values of $g_{12}$ closer to $1$ shown in 
Fig.~\ref{fig:statsolpropabc}, increasing $\mu_b$ results in a larger 
mass ratio $M$, larger FWHM $W$, and larger amplitude $A$. 
In this regime, higher mass ratio solutions are wider and taller.  
In contrast, for the relatively larger values of $g_{12}$ shown 
in Fig.~\ref{fig:statsolpropdef}, namely farther into the immiscible 
regime, it is the lower $\mu_b$ values that correspond to higher 
mass ratio $M$, and as $\mu_b$ decreases this corresponds to an 
inverse relation between $A$ and $W$ (cf.~as width increases, 
the amplitude decreases).  
In this regime, higher mass ratio solutions are wider but also shorter.  
In Fig.~\ref{fig:statsol-imagesc} we depict two typical examples of
the densities of the VB complex stationary solution, one in each of 
the above-mentioned regimes.
Figure~\ref{fig:statsol-cut} shows the corresponding density cross 
sections. Figures \ref{fig:statsol-imagesc} and \ref{fig:statsol-cut}
tend to suggest that, for the same mass ratio $M$, bright spots in the 
lower $g_{12}$ regime are tapered at the top with a wide base 
(appearing with soft edges) while bright spots in the higher $g_{12}$ 
regime appear less tapered (sharper with more well-defined boundaries). 
This is rather intuitive on account of the
deeper immiscibility region arising as
$g_{12} \gg 1$.

\begin{figure}
\includegraphics[width=1.0\columnwidth]{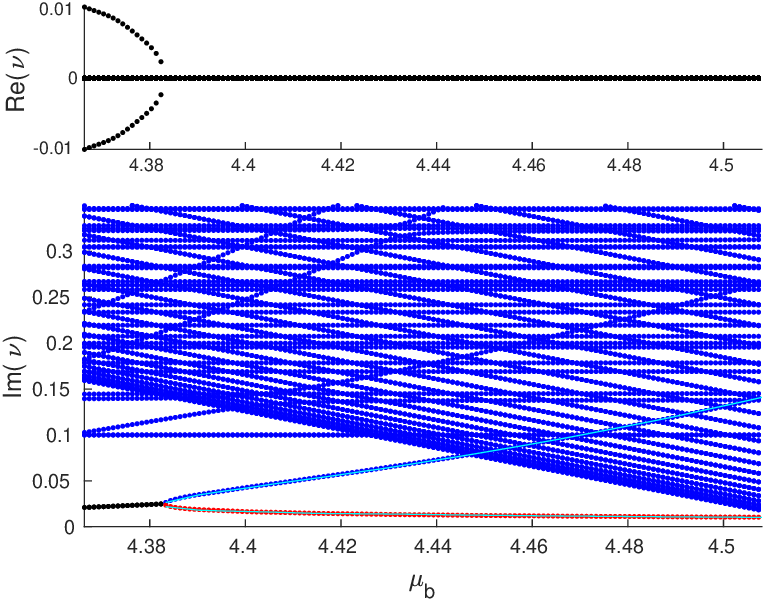}
\caption{
BdG stability spectrum for a VB structure centered at the 
origin as a function of $\mu_b$ for $g_{12}=1.6$, $\mu_d=4.25$, 
and $\Omega=0.1$.  
The top and bottom panels depict, respectively, the real and imaginary
parts of the eigenvalues. A positive real part indicates instability.
The eigenvalues are color-coded according to the sign of the Krein 
signature of the corresponding eigenvector: zero (black), negative (blue), 
and positive (red).   
The resulting bifurcation diagram evidences the presence of a 
(reverse) Hamiltonian Hopf bifurcation at $\mu_b\approx 4.383$.   
Overlaid are plots (cyan) of $\dot\theta$ values of the matching ODE 
given by Eq.~(\ref{eq:freqpm}). 
}
\label{fig:alleigs}
\end{figure}

Let us now track the BdG eigenvalue spectrum for the centered
VB structure as a function of the parameter $\mu_b$. 
Figure~\ref{fig:alleigs} depicts a typical spectrum  for
$g_{12}=1.6$ and $\mu_d=4.25$. 
Note that Re$(\nu)>0$ indicates an instability.
Interestingly, as the spectrum shows, 
there is a Hamiltonian Hopf bifurcation as $\mu_b$ decreases.  
For $\mu_d=4.25$ and values $g_{12}=1.4, 1.6, 1.75$, the bifurcations 
occur, respectively, at $\mu_b\approx 4.349, 4.383, 4.403$ and where 
according to Fig.~\ref{fig:statsolpropdef} the respective critical 
mass ratios are $M^*\approx 0.057, 0.051, 0.048$.
This (reverse) Hamiltonian Hopf bifurcation evidences that the lowest two
oscillatory (purely imaginary) eigenfrequencies collide as $\mu_b$ 
decreases giving rise to an unstable (complex) quartet. 
The presence of a complex quartet instability indicates that, in this
parametric regime, the VB will destabilize from the center of
the trap along an spiraling-out trajectory as we will see below soon.
According to the color-coded eigenvalues in the bottom panel of Fig.~\ref{fig:alleigs}, the 
region of instability lies where the lower branch is colored black 
for $\mu_b\lesssim 4.383$ and where the mass ratio $M>M^*$ exceeds 
the critical value $M^*\approx 0.051$. 
However, it is interesting that, with the parameters provided 
in Fig.~\ref{fig:alleigs}, spiraling-out is also present for $\mu_b$ 
values as high as $\mu_b<4.4$ when the VB structure is displaced by 
a large value such as $x_0=2.5$. This tends to suggest that precessional
circular orbits may also be prone to instabilities.
This is a topic of particular interest in its
own right, meriting an examination of such
states in a co-rotating frame (as was done, e.g.,
in the work of Ref.~\cite{kivpel}), yet this type of
distinct study is an interesting topic for future work.

Focusing on initially centered VB solutions, according to our BdG spectral 
analysis, starting infinitesimally close to the center, spiral-out trajectories 
will only be present in the $\mu_b$-interval where the stationary solutions 
are unstable.  
For instance, the VB dynamics for an initial condition 
$(x_0,y_0)=(2.5,0)$ in the stable regime ($\mu_b\approx 4.4$)
displays flowering trajectories that have roughly two petals and 
as $\mu_b$ increases the number of petals increase.
In Fig.~\ref{fig:alleigs}, we overlay the two frequencies obtained
from the ODE model given by Eq.~(\ref{eq:freqpm}) (see the two cyan curves).
In the stable regime, these ODE frequencies correspond to the two 
eigenfrequencies emerging from the Hamiltonian Hopf bifurcation within the PDE model.
Note the excellent agreement between these two BdG eigenfrequencies and
their ODE counterparts. 
It is important to note that this agreement is valid for a relatively 
wide range of $\mu_b$ values since the fit of the ODE frequencies, and 
hence the fit of the parameters $M_0$ and $\Lambda$ in Eqs.~(\ref{eq:ode}),
was performed for each value of $\mu_b$.
Nonetheless, we checked that, for each chosen value of $\mu_b$ (and 
the corresponding fitted parameters), the PDE and ODE orbits matched
for a wide range of initial conditions. This evidences the validity
of the ODE model not only qualitatively, but more importantly, 
quantitatively in its ability to capture the VB evolution within the full PDEs once the parameters of the system are set.

\begin{figure*}
\includegraphics[height=7.2cm]{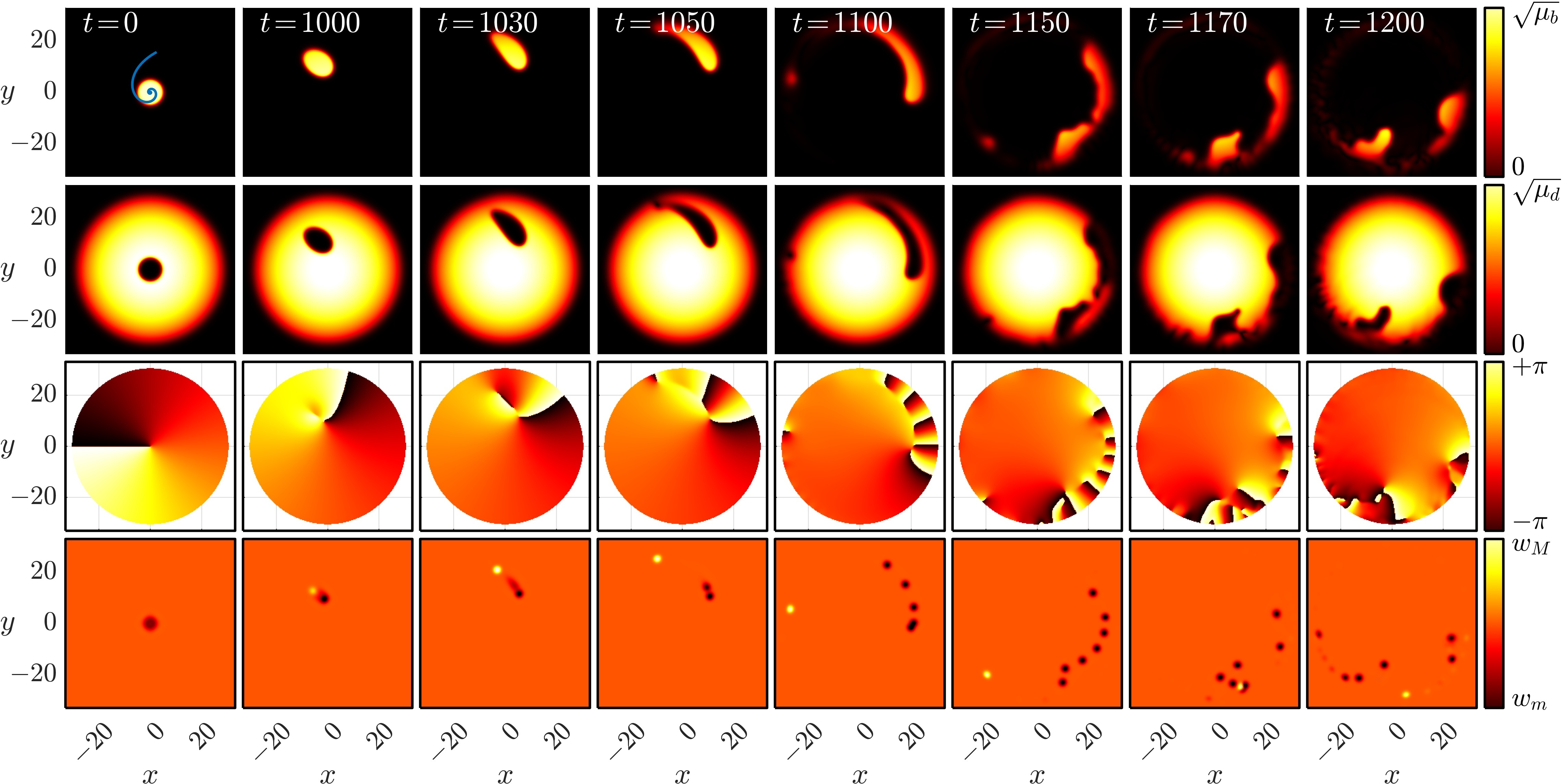}
~~
\includegraphics[height=7.2cm]{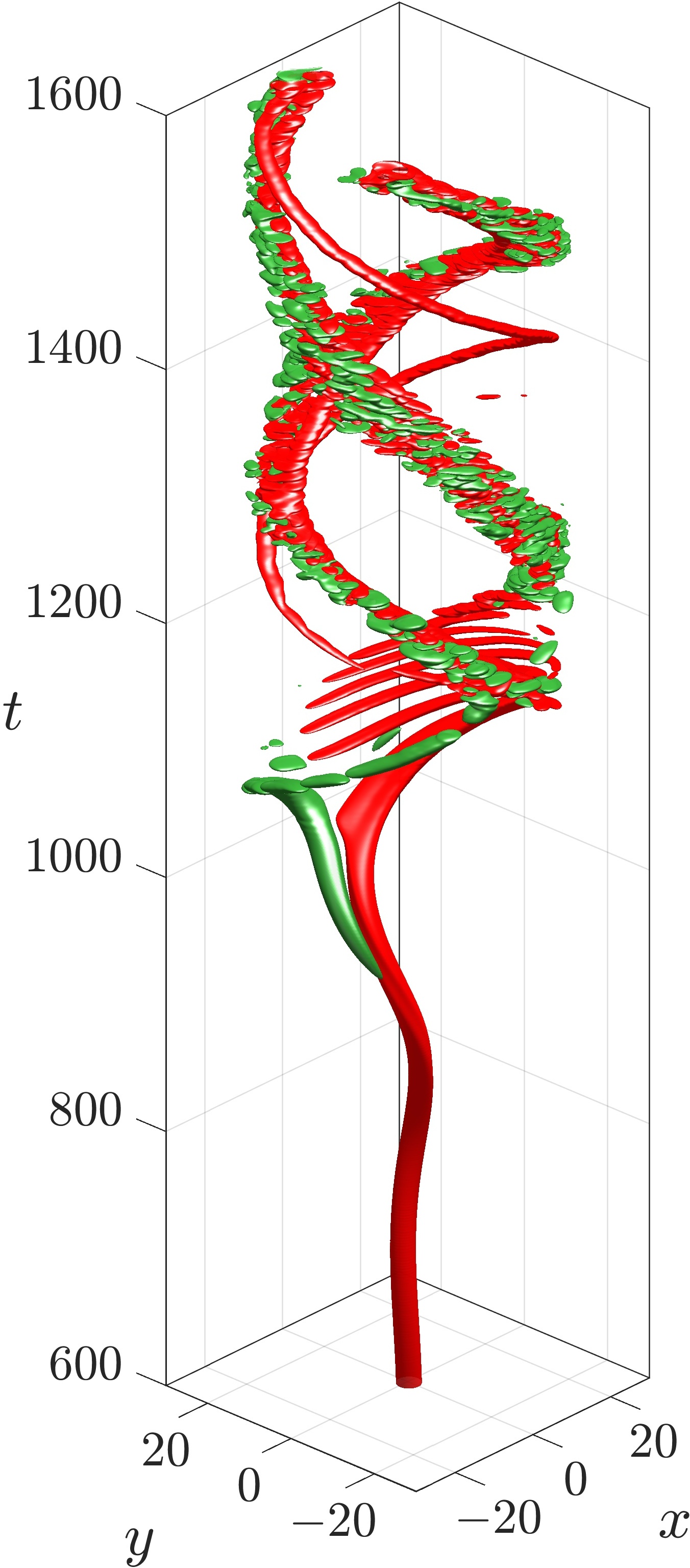}
\caption{
Evolution of an initially centered unstable VB structure
for $\mu_d=4.25$ and $\mu_b\approx4.38$. 
The first and second rows of panels depict the (square root) 
of the densities for, respectively, the bright and 
dark components at the times indicated.
The blue spiral in the top-left panel corresponds to
the orbit followed by the VB structure during
its destabilization for $0\leq t \leq 1000$.
The third and fourth rows of panels depict,
respectively, the phase
and vorticity of the dark component.
For convenience, the phase panels only show the 
region inside the Thomas-Fermi radius (outside this 
region the phases are very complicated as the 
density rapidly decays to zero).
The yellow (black) dots in the vorticity panels 
depict the location of positively (negatively) 
charged vortices where the vorticity attains 
its maximum (minimum) denoted by $w_M$ ($w_m$).
The right panel depicts the evolution of the 
vorticity by means of isocontours of vorticity
in $(x,y,t)$ where the red and green contours correspond
to negative and positive vorticities, respectively.
}
\label{fig:spiral_out_plots}
\end{figure*}

On the other hand, for our chosen parameters and as per our stability analysis, 
for $\mu_b<4.4$ the VB structure centered at the origin is unstable and tends to 
spiral out.
We provide a typical example of the destabilization dynamics
in this region in Fig.~\ref{fig:spiral_out_plots} which depicts the 
time-evolved solution together with its phase and vorticity profiles.  
As expected, for initial times ($t\lesssim 1000$), the VB destabilizes from
the origin and performs a spiraling-out motion (see overlaid blue curve 
in the top-left panel) while, approximately, preserving its VB initial 
shape albeit a small elongation along the spiraling direction.
Further down the evolution, this elongation of the VB ``patch'' becomes more 
pronounced allowing for the formation of extra vortices inside  the 
dark component of the VB. 
Initially, a vortex-anti-vortex pair creation occurs inside 
the VB complex (see the state at $t=1030$). 
Then, the corresponding negatively charged vortex keeps trapped
within the VB complex while the positively charged vortex, 
carrying a small amount of bright mass, is ejected out of the 
VB and runs toward the edge of the condensate orbiting in the 
opposite direction of the VB motion.
Next, as the VB patch continues to stretch, a series of aligned 
vortices of negative charge are created giving rise to a
quasi-1D vortex ``patch''. Eventually, the stretched VB 
patch breaks into two smaller VB patches, each carrying 
a portion of the vortices.
We note that, to the best of our knowledge, this intrinsic creation
of vortices (i.e., without the need of an external stirrer or potential) 
is associated with a novel type of VB-creating instability. 
What seems to be happening here involves the transfer
of angular momentum associated with the spiraling
motion of the VB into net vorticity since the vortices 
produced are predominantly of the same charge.
This is partially reminiscent of the process of generation of 
persistent currents (see, e.g., Ref.~\cite{persistent}). A slightly
different in spirit, yet also intriguing, recent example of vortical
phase structure produced in the bright components of VB complexes in rotating
traps, can be found in~Ref.\cite{yakimenko}.
The latter so-called ghost vortices can be rendered real through suitable
interaction quench protocols as discussed in Ref.~\cite{yakimenko}.

\begin{figure}
\includegraphics[width=0.9\columnwidth]{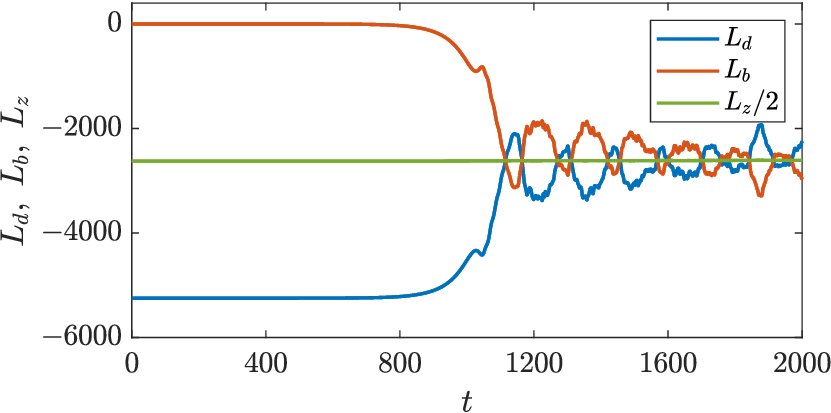}
\caption{
Temporal evolution of the angular momentum for the
destabilization dynamics of an initially centered
VB depicted in Fig.~\ref{fig:spiral_out_plots}.
The initial gradual angular momentum exchange starting around
$t\approx 800$ corresponds to the spiraling-out of the VB.
As time progresses, the angular momentum of each component tends to 
vary around the equipartition level $L_d\approx L_b\approx L_z/2$. 
}
\label{fig:Lz}
\end{figure}

In order to garner more information about the emergence of new
vortices through the above-mentioned destabilization dynamics of the
centered VB complex, we monitor the vorticity in the system.
In particular, the total angular momentum in each component can be computed 
using~\cite{Mukherjee2020}:
\begin{equation}
\notag
L_{d,b}(t)=-i\iint_{-\infty}^{+\infty} \Psi^*_{d.b}\left(x\partial _{y}-y\partial_{x}\right)\Psi_{d,b}\,dx\,dy,
\end{equation}
where the subscripts denote the dark and
bright components. Since the underlying potentials
for both components are independent of the angular coordinate,
the total angular momentum of the system is conserved:
\begin{equation}
\notag
L_z(t)=L_d(t)+L_b(t)=L_z(0).
\end{equation}
Figure~\ref{fig:Lz} illustrates the evolution of the angular
momentum for the destabilization dynamics of the initially
centered VB complex described in Fig.~\ref{fig:spiral_out_plots}.
As we saw in Fig.~\ref{fig:spiral_out_plots}, the VB
complex evolves and creates vortices of different charges
during its evolution. However, as Fig.~\ref{fig:Lz} confirms,
the total angular momentum is indeed conserved.
It is interesting to note that the angular momentum transfer 
from the dark to the bright component seems to occur as a
result of the spiral-out motion (which only starts to become 
significant after $t=800$). 
In this process, the bright soliton gains angular speed and 
moves further away from the origin as thus increasing $|L_b|$
at the expense of $|L_d|$.

Subsequently, as vortex nucleations and annihilations ensue, there
is a back-and-forth exchange of angular momentum between the two
components. Nonetheless, it is particularly interesting to note 
that in this case the system tends to ``equilibrate'' towards a 
state that has relatively small angular momentum exchanges between
the components that hover around the mean value of the total
angular momentum such that $L_d\approx L_b\approx L_z/2$.
It would be interesting to study this apparent angular 
momentum equipartition between the components in systems 
with large number of vortices in the quantum turbulent regime. 
It is also relevant to point out that in Fig.~\ref{fig:spiral_out_plots}, 
the majority of the vortices created feature a circulation 
in the negative direction. This is natural to expect
given the selection of a spiraling direction of the
vorticity patch upon the manifestation of the 
Hamiltonian Hopf bifurcation-related instability.
In addition, this spiral direction is a result of the negatively 
charged vortex state, i.e., it is the same as the precessional 
direction of the initial VB state in the harmonic trap
and can be reversed for vortices of opposite charge.

\subsection{Vortex-bright dynamics}
\label{sec:VB_Dyn}

We now leverage the ODE reduction dynamics presented in Sec.~\ref{sec:ODE}
to describe the dynamics of the VB in the original PDE model.
To generate rotational dynamics of the VB state at the level of the
original GP model~(\ref{eq:GPE}), we displace the stationary solutions 
away from the origin and seed the waveform into a ground state cloud 
according to the following procedure. First we compute, using our
fixed-point algorithm, the ground state cloud $\psi_1=\psi_{\rm gs}$ 
for a given chemical potential value $\mu_d$ and a fixed value of
the inter-species coupling $g_{12}$ (while we keep the trap strength
constant $\Omega=0.1$ throughout this analysis). For convenience we choose 
a starting point with $x(0)=x_0$ and $y(0)=0$. 
Next, we evaluate the local chemical potential 
$\mu_{x_0}=|\psi_{\rm gs}(x_0,0)|^2$ where the VB will be seeded. 
The VB solution with $\mu_d=\mu_{x_0}$ 
for a chosen $\mu_b$ value is prepared and subsequently imprinted at 
the location $x_0$ onto the ground state cloud $\psi_{\rm gs}$.   
The resulting displaced structure is propagated according to the 
dynamical equation~(\ref{eq:GPE}). Note that to avoid unnecessary 
oscillations of the background cloud induced by the 
imprinted VB, the initial condition is ``kicked'' with a suitable
(small) velocity $k$ by boosting $\psi_2$ with $e^{ik y}$ in 
order to eliminate any linear momentum for the dark cloud.
Our study of the resulting VB trajectories will focus mostly 
on parameter regimes where flowering trajectories are obtained.  

For the relatively small $g_{12}$ values shown in Fig.~\ref{fig:statsolpropabc}, 
the resulting PDE trajectories after displacement away from the origin are 
observed to be approximately circular with additional wobbling motion
that does not genuinely resemble flowering trajectories.
Nonetheless, for the relatively larger $g_{12}$ values shown 
in Fig.~\ref{fig:statsolpropdef} and for large enough $\mu_b$ values,
the PDE trajectories result in flowering motion. It is in this regime that
we will use the ODE reduction provided in Sec.~\ref{sec:thy} to match
the full GP VB dynamics.   
For example, keeping $\Omega = 0.1$ and for $g_{12}=1.6$, $\mu_d=4.25$, 
and $x_0=2.5$, GP flowering trajectories are seen for $\mu_b \gtrsim 4.4$ 
while spiral-outs are seen for $\mu_b < 4.4$. 
This is consonant with the stability results depicted in Fig.~\ref{fig:alleigs}.
However, it is important to note that the stability results of 
Fig.~\ref{fig:alleigs} correspond to a VB that is {\em centered at the origin}
while we are now seeding VB structures away from the center. 
For instance, we note that within the $\mu_b$-interval $4.383<\mu_b<4.4$ 
stationary (i.e., centered at the origin) solutions are stable, but the 
displaced state starting at $x_0=2.5$ results in a spiral-out motion. 
This evidences that this displacement is relatively far away from the 
origin for the (linear) stability analysis at the origin to be valid.  
Nonetheless, we have confirmed that, in the unstable region 
$\mu_b\lesssim 4.383$, slightly perturbed solutions from the 
origin tend to oscillate with frequencies given by Im$(\nu)$.

\begin{figure}
\centerline{
\includegraphics[width=0.9\columnwidth]{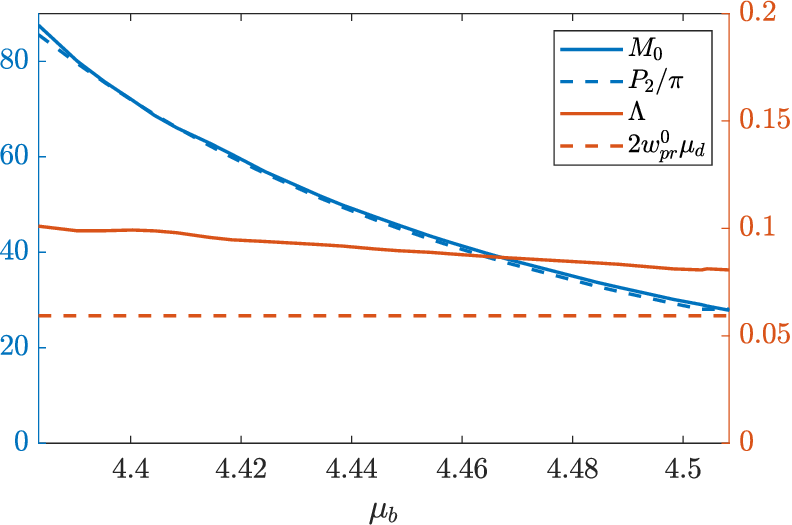}
}
\caption{
Plot of the ODE parameters $\Lambda$ and $M_0$
that best match the PDE trajectories with zero initial velocity. 
Same parameters as in Fig.~\ref{fig:alleigs}. 
The blue dashed line corresponds to the estimate of the parameter $M_0$ 
from Ref.~\cite{ruban22} and the red dashed line shows the known zero-bright-mass 
frequency $\Lambda\approx 2\omega_{\rm pr}^0\mu_d$ (see text for details). 
}
\label{fig:ODEparams}
\end{figure}

To compare the ODE reduction with the full GP dynamics in the stable regime of 
Fig.~\ref{fig:alleigs} we also depicted therein the value for $\dot\theta$ 
from the ODE model (see the two cyan curves) overlaid to demonstrate 
the frequency matching between the PDE and ODE models. This matching is 
done according to the following procedure.
For a given flowering trajectory in the PDE model within appropriate parameter 
regimes, to obtain the corresponding ODE parameters
we first identify those eigenvalues where the Krein signature bifurcates 
from zero to positive and negative values.  
The eigenvalues in Fig.~\ref{fig:alleigs} are color coded using 
(blue, black, red) points when the Krein sign is (negative, zero, positive).
In particular, for a given $\mu_b$ value, we use the upper branch of the 
bifurcated eigenvalues to obtain the value $\dot\theta = {\rm Im}(\nu)$.  
Then, by Eq.~(\ref{eq:quadratic}) we write 
the corresponding ODE $\Lambda$ parameter as:
\begin{equation}
\Lambda = 2[\rho(r_0)\dot\theta-M_0\dot\theta^2].
\notag
\end{equation}
Having obtained $\Lambda$ as a function of $M_0$, a standard minimization procedure 
is then used to determine the initial VB ODE velocity and the $M_0$ value that best 
fit (in the least-squared sense) the ODE trajectory to the PDE trajectory.   
As can be seen in Fig.~\ref{fig:alleigs} (see cyan curves), 
this fitting procedure allows to accurately obtain the {\it entire} lower branch 
of the PDE bifurcating eigenvalues from the ODE frequencies of 
Eq.~(\ref{eq:freqpm}) as well as the critical point of the Hamiltonian Hopf bifurcation.
This very good agreement lends strong credibility to our ODE reduction 
picture of the VB structure, i.e., Eqs.~(\ref{eq:ode}).

\begin{figure}
\includegraphics[width=1.0\columnwidth]{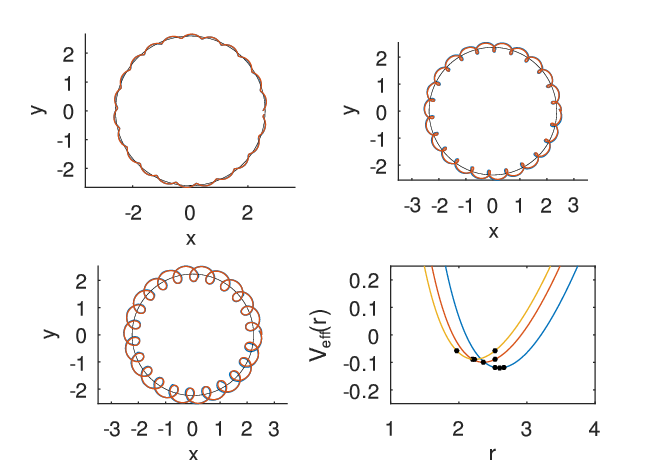}
\caption{
VB trajectories for $g_{12}=1.6$, $M=0.01$, $\mu_d=4.25$, and $\mu_b=4.56$. 
The VB is seeded at $x_0 = 2.5$ corresponding to a local
chemical potential $\mu_{x_0}\approx 4.22$.
To match the ODE trajectories we use $M_0\approx 18.9$.   
The top-left panel shows the resulting trajectory with initial velocities 
$k_y =0$ in the PDE (blue curve) and $\dot{y}(0) \approx 0.01$ in the ODE 
(red curve) with a resulting angular momentum of $\ell \approx 8.9$.  
For this value of $\ell$, the effective potential $V_{\rm eff}(r)$ in 
Eq.~(\ref{eq:Veffell}) is plotted in the bottom-right panel (blue).  
The three black dots plotted on $V_{\rm eff}(r)$ correspond to its 
minimum and the minimum and maximum radii of the trajectory.  
From these values we have an effective flower energy of $\Delta E \approx 0.002$ 
according to Eq.~(\ref{eq:deltaE}).  
Parameters are the same in the top-right panel except with initial 
velocities $k_y =0.1$ in the PDE and $\dot{y}(0) \approx 0.06$ in the ODE, 
with $\ell \approx 8.92$ giving the effective potential in the bottom-right 
panel (red) with $\Delta E \approx 0.01$.  
Similarly, the bottom-left panel corresponds to $k_y =0.15$ in the PDE
and $\dot{y}(0) \approx 0.08$ in the ODE, while $\ell\approx 8.93$  giving the effective potential in the bottom-right panel (yellow) with $\Delta E \approx 0.03$. 
The thin black circles in the trajectory plots correspond to the circular orbits
at the minima of the effective potentials.
}
\label{fig:flowers01}
\end{figure}

\begin{figure}
\includegraphics[width=1.0\columnwidth]{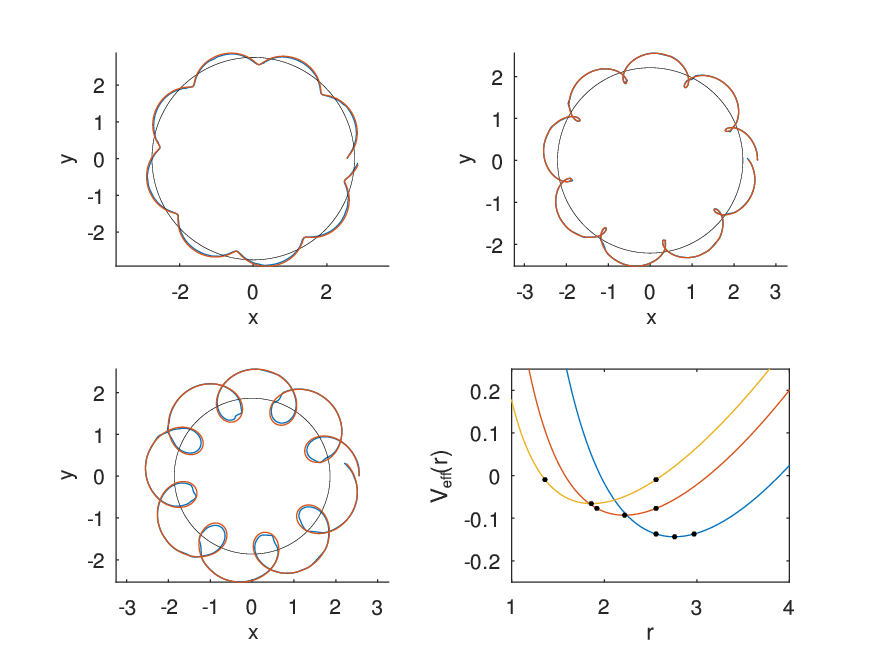}
\caption{
Similar to Fig. \ref{fig:flowers01} with the same parameters except here 
$\mu_b\approx 4.47$, $M=0.02$, and $M_0\approx 37.0$.  The top left panel 
has initial velocities $k_y =0$ in the PDE and $\dot{y}(0) \approx 0.01$ 
in the ODE, $\ell \approx 8.9$, and $\Delta E \approx 0.006$.  For this $\ell$,  
$V_{\rm eff}(r)$ is in the bottom right panel (blue). For the top right panel 
$k_y =0.1$ in the PDE, $\dot{y}(0) \approx 0.06$ in the ODE, with $\ell \approx 8.95$ 
to give the effective potential in the bottom right panel (red) with 
$\Delta E \approx 0.016$.  Similarly in the bottom left panel $k_y =0.15$ 
in the PDE, $\dot{y}(0) \approx 0.08$ in the ODE, $\ell\approx 8.97$, and 
the corresponding effective potential is in the bottom right (yellow) 
with $\Delta E \approx 0.056$. 
}
\label{fig:flowers02}
\end{figure}

Figure~\ref{fig:ODEparams} depicts the corresponding ODE parameters across 
a range of $\mu_b$ values to the right of the Hamiltonian Hopf bifurcation point.
The blue dashed line $P_2/\pi$ depicts the estimate 
for our ODE parameter $M_0$ according to Ref.~\cite{ruban22} 
[note their $(\tilde{M}, M_{br})$ corresponds to our $(M_0, P_2)$ respectively 
as their Eq.~(29) agrees with our Eq.~(\ref{eq:ode}); see also their 
Eq.~(24) giving $M_0=M_{br}/\pi=P_2/\pi$].  
Notice that our model differs from Refs.~\cite{RichaudFetter:21} 
and~\cite{richaudgeneral} where they use parameter $\mu$ to represent 
the mass ratio which is $M=P_2/P_1$ in our notation.
The match of the relevant curves in Fig.~\ref{fig:ODEparams} is very
good, revealing only  a slight relative discrepancy of no more than 
$3.5\%$ for the estimate of $M_0$.
On the other hand, the red dashed line corresponds to the precessional frequency 
of a vortex in the absence of the bright component given by
$\Lambda=2\omega_{\rm pr}^0\mu_d$ obtained from Ref.~\cite{middel} as follows.  
Taking our Eq.~(\ref{eq:ode}) with $M_0=0$ near the origin we obtain 
\begin{eqnarray}
\mu_d\, \dot{x} &=& -\frac{\Lambda}{2}y \notag\\
\mu_d\, \dot{y} &=& \frac{\Lambda}{2}x. \notag
\end{eqnarray}
Comparing this reduced system to Eqs.~(3) and (4) of Ref.~\cite{middel} 
gives $\Lambda = 2w_{\rm pr}^0\mu_{d}\equiv \Omega^2 \ln(8.88\mu_{d}/\Omega)$.
The relatively large discrepancy between $\Lambda$ and the theoretical 
estimate may stem from the fact that the latter, rather than being
a true estimate for the precessional frequency of the (massive) VB, it is just 
the precessional frequency corresponding to a vortex in the {\em absence} of the
bright component.
Note that, for this region of parameters (see Fig.~\ref{fig:statsolpropdef}),
as $\mu_b$ increases the mass of the bright component decreases and thus, 
correspondingly, the $\Lambda$ curve progressively approaches the (red dashed) 
$\Lambda=2\omega_{pr}^0\mu_d$ non-filled vortex benchmark.
In the two-component VB setting, these results render evident the dependence of 
$\Lambda$ on the bright component, suitably dressing the effective (parabolic 
confinement) landscape that the composite VB structure encounters in 
connection to its motion.

Direct comparisons between the PDE and ODE orbits are presented 
in Figs.~\ref{fig:flowers01} and \ref{fig:flowers02} where the 
blue and red curves correspond to the PDE and ODE orbits respectively
(the black circle corresponds to the circular orbit at
the effective potential minimum; see below).
The different cases in each figure are shown for a fixed value 
of $g_{12}$ and a fixed mass ratio $M$ for various initial velocities.  
For each trajectory, the angular momentum $\ell$ is obtained according 
to Eq.~(\ref{eq:angmom}) and we compute the effective energy of the flower 
defined as: 
\begin{equation}
\Delta E = V_{\rm eff}(r_{\rm max})-{\rm min}\{V_{\rm eff}\},
\label{eq:deltaE}
\end{equation}
where $r_{\rm max}$ is the maximum radius (distance to the origin) 
attained by the orbit.
In the bottom-right panels of Figs.~\ref{fig:flowers01} and \ref{fig:flowers02}
we also depict the ODE effective potential $V_{\rm eff}(r)$ according to 
Eq.~(\ref{eq:Veffell}).  
The corresponding overlaid black dots represent the minimum and extremal
values of $V_{\rm eff}(r)$ attained through the flowering motion.
For guidance we also overlay (see corresponding black circles), on top 
of the flowering trajectories, the circular orbits corresponding to 
the effective potential minima.
Note that larger values of $\Delta E$ correspond to trajectories
with bigger petals.
It is also worth mentioning that, as it can be noticed from the values 
reported in Fig.~\ref{fig:flowers01}, the initial velocities for the PDE
($k_y$) are larger than the fitted initial ODE velocities ($\dot{y}$).
This apparent discrepancy is due to the fact that the VB initialization
in the PDE is not perfect as we are just boosting the bright component
with a velocity $k_y$ while we are not adjusting the velocity of the dark
component. Therefore, part of the initial energy imparted to the VB
in the PDE case is lost to background radiation while the bright
hump accelerates the vortex in the dark component. As a result, the 
effective velocity that the VB acquires, after the initial radiation is emitted, 
is actually slightly smaller than $k_y$. Finding a better initialization
procedure for the VB in the PDE model is an
intriguing topic for further consideration. 
The associated difficulty is intimately connected
to the ``dual character'' of the VB. On the one
hand, it behaves as a vortical particle (for which
at the one-component level, effective equations
involve first-order ODEs and an initialization
based purely on position), while it also behaves
as a Newtonian particle with (bright-induced)
mass, thus necessitating an initialization bearing
both position and momentum information.

\begin{figure}
\includegraphics[width=1.0\columnwidth]{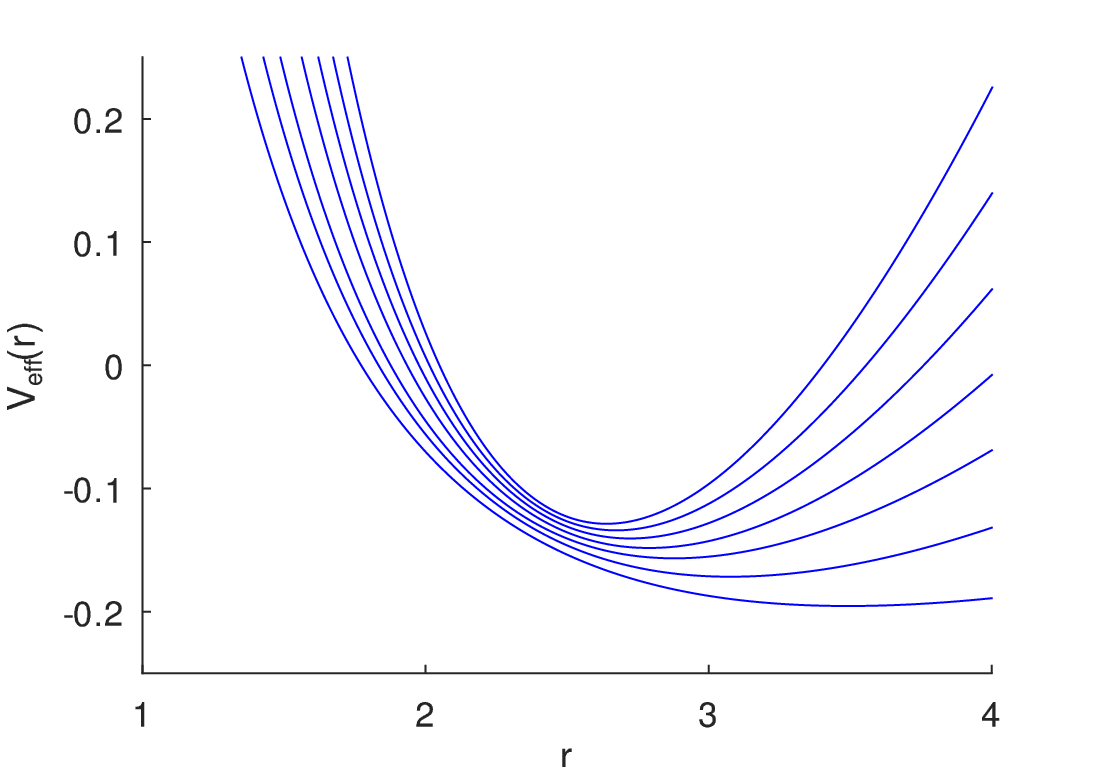}
\caption{
Effective potential $V_{\rm eff}(r)$ as the bright chemical potential 
$\mu_b$ is varied while all other parameters are the same as in 
Fig.~\ref{fig:ODEparams}. 
The different curves correspond to the values 
$\mu_b\approx 4.4, 4.42, 4.44, 4.46, 4.48, 4.5, 4.52$ where the shallower 
(steeper) effective potentials correspond to smaller (larger) values of $\mu_b$. 
}
\label{fig:Veffs}
\end{figure}

\begin{figure}
\includegraphics[width=1.0\columnwidth]{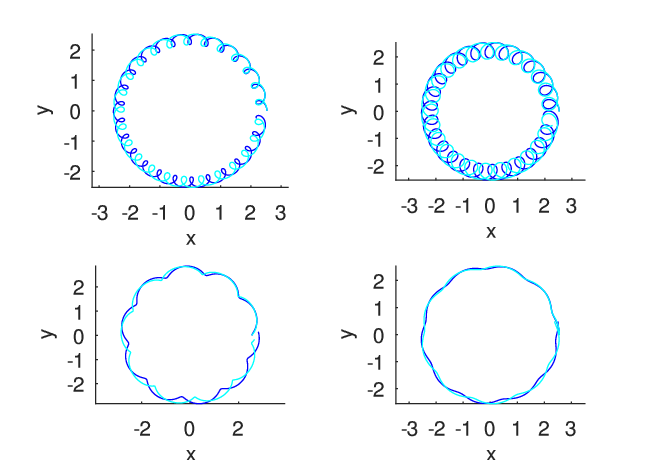}
\caption{
A comparison between flowering trajectories of the ODE system (dark blue) 
in Eqs.~(\ref{eq:ode}) and its corresponding linearization (cyan) as per
Eqs.~(\ref{eq:estRTheta}) and (\ref{eq:omegaA}).  
The top panels show the trajectories with $M_0\approx 17.95$, 
$\Lambda \approx 0.07$, and $\Omega = 0.1$.  In the top-left panel
the initial velocity is $\dot{y}(0) =0.06$ giving an
angular momentum $\ell \approx 8.93$, effective potential minimum 
at $r_0\approx 2.35$, effective energy $\Delta E \approx 0.01$, 
and the linearized prediction of $A_0\approx 0.1$, $R\approx 0.18$, 
and $(\omega_r,\omega_{\rm pr}) \approx (0.21,0.01)$.
In the top-right panel the initial velocity is $\dot{y}(0) =0.08$ 
giving $\ell \approx 8.94$, $r_0\approx 2.22$, and $\Delta E \approx 0.04$ 
with linear predicted values $A_0\approx 0.17$, $R\approx 0.31$, 
and $(\omega_r,\omega_{\rm pr}) \approx (0.21,0.01)$.
The bottom panels correspond to $M_0\approx 36.05$, $\Lambda \approx 0.08$,
$\Omega = 0.1$, and $(\omega_r,\omega_{\rm pr}) \approx (0.09,0.01)$.
The bottom-left panel corresponds to $\dot{y}(0) =0.01$, $\ell \approx 8.9$, 
$r_0\approx 2.7$, $\Delta E \approx 0.004$, $A_0\approx -0.08$, and $R\approx 0.15$.
The bottom-right panel corresponds to $\dot{y}(0) =0.03$, $\ell \approx 8.93$, 
$r_0\approx 2.49$, $\Delta E \approx 0.02$, $A_0\approx 0.04$, and $R\approx 0.07$.
}
\label{fig:odevlin}
\end{figure}

Figure~\ref{fig:Veffs} sketches several plots of the effective potential
$V_{\rm eff}(r)$ as the bright chemical potential $\mu_b$ is varied while 
all other parameters are fixed.
Since larger $\mu_b$-values correspond to smaller mass ratios $M$, the resulting 
effective potentials are steeper.  
Here, we are capitalizing on the accuracy of our ODE-PDE comparison to infer 
the dynamical consequences of parametric variation
at the PDE level, based on the ODE understanding.
As a consequence, smaller bright masses will tend to produce flowering 
with smaller petals as a steeper potential means a larger initial 
energy (initial kick) in order to produce the same petal sizes.
We should also point out that higher $g_{12}$ values result in flowering 
with a smaller frequency.    
Finally, Fig.~\ref{fig:odevlin} depicts several examples of flowering 
in the ODE and its corresponding linearized estimations as
per Eqs.~(\ref{eq:estRTheta}) and (\ref{eq:omegaA}).
Note that the linearized equations are able to capture well the
main shape and petal size of the flowering motion 
even modestly away from the linear regime. 
This suggests that Eqs.~(\ref{eq:estRTheta}) and (\ref{eq:omegaA})
can be used as simple approximate analytical descriptors of the VB
dynamics over a wide parametric range.

\section{Summary and outlook}
\label{summary}

In the present work we have  explored the topic of the dynamics
of VB solitary waves deep within the immiscible
regime. Our motivation in revisiting this subject stemmed
from the renewed experimental ability to monitor such systems
which creates the potential for connecting a particle-based
understanding with a detailed analysis of the full PDE model. 
In that vein, we explored equilibrium solutions in this
deeply immiscible regime (while some of the earlier studies
of the present authors had confined themselves to regimes
close to, e.g., hyperfine states of $^{87}$Rb where the
$g_{ij}$ coefficients are nearly equal~\cite{tuckerman}).
In probing the excitations around the VB equilibrium state,
we found the possibility of two frequencies of precession.
One of these represents the frequency of precession that
is well-familiar from a single-component vortex in a trap.
We also identified an additional frequency 
that can be associated with the oscillation of the VB structure
inside an effective radial potential (along the radial direction).
The combination of these two frequencies gives rise to the potential
for epicyclic motion, which, under suitable conditions discussed herein,
can lead to ``flowering'', i.e., to flower-like curves of the
dynamics of the VB pattern. This intriguing feature,
while observed and discussed previously~\cite{RichaudFetter:21,ruban22}
is here intimately connected between the ODE and PDE dynamics 
(both at the level of linearization and at that of fully nonlinear
dynamical evolution). 
Our analysis enables both a reasonable approximate
analytical description of the relevant motion
and key byproducts thereof, such as the critical
oscillation amplitude leading to flowering as a 
(simple) function of the precessional and radial
oscillation frequencies.
Among the additional intriguing features
identified herein, we highlight the (oscillatory) instability of the
VB structure when the relevant two frequencies collide. The
result of the corresponding dynamics was found to be particularly
unusual, in our experience of such systems. Indeed, as the VB
embarks on a growing spiral oscillation, it becomes a patch of vorticity
that gets elongated and eventually breaks into smaller patches 
carrying multiple localized phase
jumps inside an extended region of vanishing
density (in the so-called dark component)
in a way reminiscent of persistent currents~\cite{gretchen}.

While we considered a number of diagnostics (such as angular momentum-based ones) 
and visualizations (such as the 3D isocontours of vorticity)
of this phenomenon, clearly it merits further study. An additional
natural theme of study is to incorporate the (two or more) VB interactions
to the model systematically developed herein, so as to
examine the potential of formation of lattices (crystals) of such
patterns. Understanding the modes of such crystals, their potential
instabilities and dynamics promises to be an interesting avenue for the future. 
It is relevant to also point out here that an alternative formulation
of such single and multi-VB configurations is given in Ref.~\cite{richaudmultiple}. 
A comparative analysis of the latter with the effective particle system produced herein
(and its multi-VB generalizations) would be a worthwhile topic for further investigation.
Furthermore, the states of interest can be generalized to three spatial 
dimensions where the natural analogue is the vortex-ring-bright (filled core) 
solitary wave~\cite{ruban22b}; of course,
vortex-line-bright structures exist too~\cite{ksaderfos}.
A systematic particle-based understanding of such 3D settings is a natural
generalization of the considerations provided herein. Relevant
studies are currently in progress and will be reported in future publications.

\acknowledgments{
We are thankful to Victor Ruban for useful discussions and insights.
We acknowledge the support from the US National Science Foundation under
Grants No.~PHY-2110038 and No. PHY-2408989 (R.C.G.), No.~PHY-2110030,
PHY-2408988 and DMS-2204702 (P.G.K.). P.G.K. is also grateful to Profs. Vassilis Koukouloyannis and Ioannis Gkolias for discussions.
W.W. acknowledges support from the National Science Foundation of 
China under Grant No. 12004268.
}


\end{document}